\documentclass[aip, jcp, reprint]{revtex4-2}
\usepackage{graphicx}
\usepackage{multirow}
\usepackage{amsmath, bm}
\usepackage{amssymb}
\usepackage{mathtools}
\usepackage{xcolor}
\usepackage[ruled,linesnumbered,vlined]{algorithm2e}
\SetAlgoCaptionSeparator{.}
\SetAlCapHSkip{0pt}
\DontPrintSemicolon
\SetStartEndCondition{ }{}{}%
\SetKwProg{Fn}{def}{\string:}{}
\SetKwFunction{Range}{range}
\SetKw{KwTo}{in}\SetKwFor{For}{for}{\string:}{}%
\SetKwIF{If}{ElseIf}{Else}{if}{:}{elif}{else:}{}%
\SetKwFor{While}{while}{:}{fintq}%
\SetKwFunction{Append}{.append}
\SetKwFunction{Enumerate}{enumerate}
\SetKwFunction{JKkernel}{jk\_kernel}
\SetKwFunction{Rysroot}{rys\_roots<$N$>}
\SetKwFunction{RR}{rr<$N$>}
\SetKwFunction{RRip}{rr1<$N$>}
\SetKwFunction{AtomicAdd}{atomicAdd}
\SetKwComment{Comment}{\#}{}
\newcommand{\var}{\texttt}

\makeatletter
\newcommand{\AlgoCaptionFormat}{}
\newcommand{\SetAlgoCaptionFormat}[1]{\def\AlgoCaptionFormat{#1}}
\renewcommand{\algocf@makecaption@ruled}[2]{%
    \global\sbox\algocf@capbox{\hskip\AlCapHSkip%
        \setlength{\hsize}{\columnwidth}
        \addtolength{\hsize}{-2\AlCapHSkip}
        \vtop{\AlgoCaptionFormat\algocf@captiontext{#1}{#2}}}
}%
\makeatother

\SetAlgoCaptionFormat{\raggedright}
\usepackage{comment}

\DeclarePairedDelimiter\floor{\lfloor}{\rfloor}

\begin{document}


\title{Introducing GPU-acceleration into the Python-based Simulations of Chemistry Framework}

 \author{Rui Li}
 \affiliation{Division of Chemistry and Chemical Engineering, California Institute of Technology}

 \author{Qiming Sun}
 \affiliation{Quantum Engine LLC.}


 \author{Xing Zhang}
 \affiliation{Division of Chemistry and Chemical Engineering, California Institute of Technology}

 \author{Garnet Kin-Lic Chan}
 \email{gkc1000@gmail.com}
 \affiliation{Division of Chemistry and Chemical Engineering, California Institute of Technology}

\begin{abstract}
We introduce the first version of \textsc{GPU4PySCF}, a module that provides GPU acceleration of methods in \textsc{PySCF}.
As a core functionality, this provides a GPU implementation of two-electron repulsion integrals (ERIs) for contracted basis sets comprising up to $g$ functions using Rys quadrature. As an illustration of how this can accelerate a quantum chemistry workflow, we describe how to use the ERIs efficiently in the integral-direct Hartree-Fock Fock build and nuclear gradient construction.
Benchmark calculations show a significant speedup of two orders of magnitude
with respect to the multi-threaded CPU Hartree-Fock code of \textsc{PySCF},
and performance comparable to other GPU-accelerated quantum chemical packages
including \textsc{GAMESS} and \textsc{QUICK} on a single NVIDIA A100 GPU.
\end{abstract}

\maketitle


\section{Introduction}
The rapid advances in the capabilities of graphics processing units (GPUs) has significantly impacted many fields,
including graphics rendering, gaming, and artificial intelligence.\cite{NEURIPS2019_9015,tensorflow}
The massively parallel architecture of GPUs offers drastically more computational throughput than
traditional central processing units (CPUs),
making them well-suited for computationally intensive tasks such as dense matrix multiplication and tensor contraction.\cite{springer2019cutensor}
Consequently, GPUs have evolved into powerful tools for scientific computation
on high-performance computing (HPC) platforms.
For instance, at the National Energy Research Scientific Computing Center, GPUs deliver a maximum compute performance of 119.8 PFLOPS
compared to only 11.6 PFLOPS from the associated CPUs.\cite{NERSCArchitecture}
However, leveraging GPUs for substantial performance gains over CPUs typically requires significant redesign of algorithms.

In the field of quantum chemistry, GPUs have been extensively explored to accelerate the Hartree-Fock (HF) and density functional theory (DFT) methods. Particular attention has been paid to evaluating two-electron repulsion integrals (ERIs), a key computational primitive, and their  subsequent use in the Fock builds of the HF and DFT equations.
Over the past 15 years, various GPU algorithms for ERI evaluation and the Fock builds have been proposed.
Yasuda\cite{Yasuda2008} implemented the first such algorithm,
along with the construction of the Coulomb matrix
using the J engine method.\cite{White1996,Challacombe1997}
At the same time, Ufimtsev and Mart{\'i}nez\cite{Ufimtsev2008,Ufimtsev2009} developed a GPU implementation for the HF method, which included building the full Fock matrix
(both Coulomb and exchange matrices), with ERIs evaluated using the McMurchie-Davidson (MD) algorithm.\cite{McMurchie1978}
Both implementations were initially limited to Gaussian basis sets containing only $s$ and $p$ functions.
Later, Asadchev and Gordon\cite{Asadchev2010}
developed a Fock build algorithm using the Rys quadrature method\cite{Dupuis1976, Rys1983} for ERI evaluation, allowing for the use of uncontracted basis sets with up to $g$ functions.
In addition, Miao and Merz\cite{Miao2013} employed the Head-Gordon-Pople (HGP) algorithm\cite{MHG2e}
to reduce the number of floating-point operations (FLOPs) required for computing ERIs
with contracted basis functions.
Recently, Barca \textit{et al.} introduced a distinct implementation of the HGP algorithm
and an improved ERI digestion (the contraction between ERIs and the density matrix to form the Fock matrix) scheme.\cite{Barca2020}
This was subsequently extended to run on multiple GPUs.\cite{Barca2021,gamessLatest}
Their code outperformed most previous multi-GPU implementations,
but significant performance drops were observed for ERIs involving basis functions with higher angular momenta, such as $d$ functions.
To address this issue,
Asadchev and Valeev\cite{AsadchevFourCenter,Asadchev2024}
developed a matrix-based formulation of the MD algorithm,
leveraging extensive use of dense matrix multiplication kernels.
Their approach achieved significant speedups over the reference CPU implementation,
particularly for high angular momentum ERIs, including those involving $i$ functions.

In this work, we describe our implementation of four-center ERIs on the GPU within the \textsc{GPU4PySCF} module. As a core computational routine, this was the first feature to be developed. At the time of writing, \textsc{GPU4PySCF} also contains many additional features, including those developed using GPU-accelerated density fitting ERIs~\cite{Wu2024} (which are computed using an adaptation of the four-center ERI algorithm). However, to limit the scope of this paper to the work of the current set of authors, as well as to present the chronological development of the package, this work describes only the algorithm for four-center ERIs and the subsequent Fock build routines that use them.

Our ERI implementation is based on
Rys quadrature.
One advantage of this technique is that it features a small memory footprint,
making it well-suited for mainstream commodity GPUs with limited fast on-chip memory.
Additionally, it offers simple recurrence relations,
facilitating straightforward extensions to high angular momentum and ERI derivatives.
Within this framework, we utilize several algorithmic optimizations to enhance the performance of both energy and nuclear gradient
ERI evaluation
with the latest compute unified device architecture (CUDA). The resulting ERI routines
support contracted basis sets comprising up to
$g$ functions.


\textsc{GPU4PySCF} is designed to operate primarily within the Python environment.
Consequently, in addition to the custom CUDA kernels for the ERIs, it utilizes \textsc{NumPy}\cite{numpy}-like packages (such as \textsc{CuPy}\cite{cupy_learningsys2017}) to accelerate the
computationally expensive tensor contractions and linear algebra operations on the GPU.
The Python-based nature of \textsc{GPU4PySCF} allows for seamless integration with
other Python-based workflows, particularly those in machine learning applications. We envision that this choice of ecosystem for quantum chemistry GPU acceleration will allow \textsc{GPU4PySCF} to achieve the same type of interdisciplinary impact that its parent package \textsc{PySCF} has become known for.\cite{Sun2020}

The paper is organized as follows.
Section \ref{sec:rys} provides a brief review of the Rys quadrature method.
Sections \ref{sec:fock} and \ref{sec:grad} detail our GPU-accelerated Hartree-Fock (HF) implementation,
focusing on the algorithms for Fock build and nuclear gradients, respectively.
The performance of our method is examined in Section \ref{sec:Result}.
Finally, we draw some conclusions and describe our general outlook for \textsc{GPU4PySCF} in Section \ref{sec:Conclusion}.

\section{Rys quadrature method}\label{sec:rys}
In the Rys quadrature method,\cite{Dupuis1976,Rys1983,Flocke2008} the six-dimensional ERI
is expressed as a product of three two-dimensional (2D) integrals ($I_x$, $I_y$, and $I_z$),
evaluated exactly by an $N$-point Gaussian quadrature with weights ($w_n$) and roots ($t_n$)
of the Rys polynomial:
\begin{equation}
    [ab|cd] = \sum_n^N w_n  I_x(t_n) I_y(t_n) I_z(t_n) \;.
    \label{eq:eri_rys}
\end{equation}
In Eq.~\eqref{eq:eri_rys}, the ERI is computed for Cartesian primitive Gaussian functions (PGFs),
\begin{equation}
    |a] \equiv \phi_a(\mathbf{r}) =
    (x-A_x)^{a_x} (y-A_y)^{a_y} (z-A_z)^{a_z} e^{- \alpha |\mathbf{r} - \mathbf{A}|^2},
\end{equation}
which are centered at nuclear positions $\mathbf{A}=(A_x, A_y, A_z)$,
with exponents $\alpha$ and angular momenta $a = a_x + a_y + a_z$.
The number of quadrature points is related to the angular momenta of the four PGFs as
\begin{equation}
    N = \floor*{\frac{a + b + c + d}{2}} + 1 \;.
\end{equation}
The 2D integrals $I_x$, $I_y$, and $I_z$ are evaluated for each primitive shell quartet
(denoted as $[\mathbf{ab}|\mathbf{cd}]$, where bold letters indicate a shell of basis functions),
using the recurrence and transfer relations (RRs).\cite{Dupuis1976,Rys1983}
Each of the 2D integral tensors has a size of $(a+1)(b+1)(c+1)(d+1)$ for each quadrature point.
Finally, for contracted Gaussian functions (CGFs),
which are linear combinations (with contraction order $K$) of PGFs,
\begin{equation}
    |i) = \sum_a^K C_a^i |a] \;,
\end{equation}
the contracted ERI can be written as
\begin{equation}
    (ij|kl) = \sum_{abcd} C_a^i C_b^j C_c^k C_d^l [ab|cd] \;.
\end{equation}

Modern GPUs offer high computational throughput but often suffer from significant memory latency.
They are well-suited for tasks with high arithmetic intensity
[defined as the ratio of FLOPs to data movement (in bytes)],
such as dense matrix multiplications, where the latency can be effectively masked.
ERI evaluation using the Rys quadrature method may or may not fall into this category of tasks
depending on the feasibility of data caching in fast memory.
We can roughly estimate its arithmetic intensity by considering
the most computationally expensive step, i.e., Eq.~\eqref{eq:eri_rys}.
If no data is cached, the arithmetic intensity is approximately $\frac{3}{16} \text{FLOP/byte}$.
This intensity is significantly below the peak FLOP rate to memory bandwidth ratio of
$\frac{9.7 \text{TFLOP/s}}{1.6 \text{TB/s}}=6.1 \text{FLOP/byte}$
for the NVIDIA A100 GPU used in this work.
According to the Roofline model,\cite{Roofline}
it suggests that the corresponding implementation will be memory-bound and likely inefficient.
On the other hand, if the ERIs and 2D integrals can be cached completely,
the arithmetic intensity becomes $\frac{3NK}{8} \text{FLOP/byte}$
(assuming we are storing the results in slow memory).
This value can be higher than the previous ratio
for large $N$ and $K$, indicating a compute-bound character.
However, GPUs typically have limited fast memory (e.g., registers and shared memory),
and hence require careful algorithmic design to achieve optimal performance.

The Rys quadrature method features a low memory footprint\cite{Asadchev2010} and high data locality,\cite{Sun2015} which allows for more effective data caching.
For instance, to compute ERIs with $N\leqslant3$, the necessary data can almost entirely fit into the registers.
Specifically, for the integral class $\mathbf{(pp|pp)}$,
the required intermediates (only considering the contracted ERIs and 2D integrals)
amount to $3^4+3\times2^4=129$ FP64 words (equivalent to 258 FP32 words).
This is just above the maximum register file size allowed for an NVIDIA GPU thread,
which is 255 FP32 words (for the microarchitectures since Kepler).
However, for larger $N$ values, the data size will inevitably exceed the available resources of fast memory. As a result, the implementation must
minimize access to slow memory (e.g., local and global memory).
In practice, we incorporate the following designs:
\begin{enumerate}
    \item For ERIs with small $N$ values, the RRs are unrolled
    to fully utilize registers.
    \item In general cases, we perform ERI digestion before contraction
    and introduce novel intermediates to minimize global memory access.
    \item When computing the nuclear gradients, double contractions
    between the ERI gradients and the density matrix are performed to directly
   obtain the energy gradients, thereby avoiding the need to store the Fock matrix gradients.
\end{enumerate}
These are detailed in Secs.~\ref{sec:fock} and \ref{sec:grad}.

\section{Fock build}\label{sec:fock}
Algorithm~\ref{alg:fock} illustrates the workflow for building the Fock matrix.
(The actual implementation incorporates vectorization
and accounts for the eight-fold permutation symmetry of the ERIs.)
Strategies similar to those employed by Barca \textit{et al.}\cite{Barca2020}
are used for integral screening and workload partitioning.
The algorithm starts by grouping the shells of CGFs that share the same
angular momentum and contraction order,
forming sets of shells denoted as $|\mathbf{a}, K_a\}$ (line 1).
Shell pairs are then constructed using Cartesian products between the shells in each group, resulting in
$|\mathbf{ab}, K_aK_b\}= |\mathbf{a}, K_a\} \otimes |\mathbf{b}, K_b\}$ (line 6).
These pairs are further ``binned'' into $n_{\mathbf{ab}}$ batches indexed by the size parameter
$s_{\mathbf{ab}}$ (line 8), defined as
\begin{equation}
    s_{\mathbf{ab}} = \begin{cases}
        \begin{aligned}
            &\floor*{\frac{\log_{10} I_{\mathbf{ab}}}{\log_{10} \tau} n_{\mathbf{ab}}}, & I_{\mathbf{ab}} < 1 \\
            &0,  & I_{\mathbf{ab}} \geqslant 1
        \end{aligned}
    \end{cases} \;,
    \label{eq:sab}
\end{equation}
where the labels for contraction orders are omitted for clarity (similarly hereafter).
In Eq.~\eqref{eq:sab}, $\tau$ is a positive integral accuracy threshold smaller than one,
$n_{\mathbf{ab}}$ is a heuristic parameter determined such that each bin contains roughly 128 shell pairs, and
\begin{equation}
    I_{\mathbf{ab}} = \max_{a \in \mathbf{a}, b \in \mathbf{b}}|(ab | ab)|^{\frac{1}{2}}
\end{equation}
is the conventional Cauchy-Schwarz bound factor.\cite{Whitten1973}
Note that the shell pairs are prescreened based on the condition
$I_{\mathbf{ab}} > \tau$ before the binning process (line 7),
ensuring that at most $n_{\mathbf{ab}}$ bins are generated.

The main computational loops are executed over the sets of ``bra-ket'' shell quartets, namely,
$\{\mathbf{ab|cd}\} = \{\mathbf{ab}| \otimes |\mathbf{cd}\}$ (lines 10--11).
For each set, a loop over the $n_{\mathbf{cd}}$ batches of
the ket shell pairs is further carried out (line 12), within which the significant bra shell pairs are selected
according to the criterion $I_{|\mathbf{ab}\}} I_{|\mathbf{cd}\}} P_{\{\mathbf{ab|cd}\}} > \tau$ (line 14), where
\begin{equation}
    I_{|\mathbf{ab}\}} = \max_{\mathbf{a \in |a\}}, \mathbf{b \in |b\}}} I_{\mathbf{ab}} \;,
\end{equation}
and
\begin{equation}
    P_{\{\mathbf{ab|cd}\}} =
    \max_{\substack{\mathbf{i, j} \in \{\mathbf{a,b,c,d}\} \\ \mathbf{i \neq j}}} |P_{\{\mathbf{i|j}\}}|
\end{equation}
is the maximum element across the corresponding sub-blocks of the density matrix
(e.g., $P_{\{\mathbf{a|b}\}}$ represents the blocks with bra and ket basis functions belonging to $|\mathbf{a}\}$ and $|\mathbf{b}\}$ shell batches, respectively).
This screening procedure is performed at the level of batches of shell pairs,
which preserves the continuous layout of the shell pair data and
facilitates efficient coalesced memory access by the GPU threads.
Finally, the GPU kernel (\texttt{jk\_kernel}) is dispatched to compute the
Coulomb ($\mathbf{J}$) and exchange ($\mathbf{K}$) matrices
using the screened shell quartets (line 16).

\begin{algorithm}
    \var{shl\_sets} = $[|\mathbf{a}\}, |\mathbf{b}\}, \ldots]$\;
    \var{shlpr\_sets} = $[~]$\;
    \For{$|\mathbf{a}\}$ \KwTo \textup{\var{shl\_sets}}}{
        \For{$|\mathbf{b}\}$ \KwTo \textup{\var{shl\_sets}}}{
            $|\mathbf{ab}\} = [[~],]*n_{\mathbf{ab}}$\;
            \For{$|\mathbf{ab})$ \KwTo $|\mathbf{a}\} \otimes |\mathbf{b}\}$}{
                \If{$I_\mathbf{ab}>\tau$}{
                    $|\mathbf{ab}\} [s_{\mathbf{ab}}]$\Append{$|\mathbf{ab})$}\;
                }
            }
            \var{shlpr\_sets}\Append{$|\mathbf{ab}\}$}\;
        }
    }
    \For{$|\mathbf{ab}\}$ \KwTo \textup{\var{shlpr\_sets}}}{
        \For{$|\mathbf{cd}\}$ \KwTo \textup{\var{shlpr\_sets}}}{
            \For{$s_{\mathbf{cd}}$ \KwTo \Range{$n_{\mathbf{cd}}$}}{
                \For{$s_{\mathbf{ab}}$ \KwTo \Range{$n_{\mathbf{ab}}$}}{
                    \If{$I_{|\mathbf{ab}\}}[s_{\mathbf{ab}}]I_{|\mathbf{cd}\}}[s_{\mathbf{cd}}]P_{\{\mathbf{ab|cd}\}}<\tau$}{
                        break\;
                    }
                }
                \JKkernel{$|\mathbf{ab}\}[:s_{\mathbf{ab}}] \otimes |\mathbf{cd}\}[s_{\mathbf{cd}}]$} \Comment*[h]{on GPU}\;
            }
        }
    }
    \caption{Workflow of building the Fock matrix}
    \label{alg:fock}
\end{algorithm}

The Fock build is parallelized over a 2D GPU thread grid,
with the bra shell pairs distributed in one dimension and the ket shell pairs in the other.
Each thread evaluates the ERIs of a shell quartet, contracts them with the density matrix,
and accumulates the results into the $\mathbf{J}$/$\mathbf{K}$ matrix.
Workload balance among the threads is ensured given that the shell
quartets are of the same type (with respect to angular momenta and contraction orders).

As mentioned in Section~\ref{sec:rys}, the GPU kernel \texttt{jk\_kernel}
has two distinct designs depending on the value of $N$.
For $N \leqslant 3$ (see Algorithm~\ref{alg:jk_unroll}),
we use metaprogramming to unroll the loops involved in the evaluation of the RRs (line 6),
thereby explicitly storing the 2D integrals ($I_x$, $I_y$, and $I_z$) and other intermediates
in registers to minimize the memory latency.
In addition, the primitive ERIs are first contracted (line 7) before being digested (lines 8--13),
as sufficient registers are available to hold the contracted ERIs.
(Note that the contraction coefficients and the Rys quadrature weights
have been absorbed into $I_x$, $I_y$, and $I_z$ in Algorithm~\ref{alg:jk_unroll}.)
Similarly, loops associated with the ERI digestion are also unrolled, with
the final results accumulated into the $\mathbf{J}$/$\mathbf{K}$ matrix (stored in global memory)
using the atomic operation (\texttt{atomicAdd}),
which avoids explicit thread synchronizations.

\begin{algorithm}
    $\mathbf{(ab|cd)} = 0$\;
    \For{$|\mathbf{ab}]$ \KwTo $|\mathbf{ab})$}{
        \For{$|\mathbf{cd}]$ \KwTo $|\mathbf{cd})$}{
            $t,w$ = \Rysroot{} \Comment*[h]{Rys roots/weights}\;
            \For{$i$ \KwTo \Range{$N$}}{
                $I_x, I_y, I_z$ = \RR{$t[i], w[i]$} \Comment*[h]{unroll RRs}\;
                $\mathbf{(ab|cd)} \mathrel{+}= I_x * I_y * I_z$\;
            }
        }
    }
    \Comment*[h]{For each element of $\mathbf{J/K}$:}\;
    \AtomicAdd{$J_{ab}, (ab|cd)*P_{cd}$}\;
    \AtomicAdd{$J_{cd}, (ab|cd)*P_{ab}$}\;
    \AtomicAdd{$K_{ac}, (ab|cd)*P_{bd}$}\;
    \AtomicAdd{$K_{bd}, (ab|cd)*P_{ac}$}\;
    \AtomicAdd{$K_{ad}, (ab|cd)*P_{bc}$}\;
    \AtomicAdd{$K_{bc}, (ab|cd)*P_{ad}$}\;
    \caption{\texttt{jk\_kernel} for $N \leqslant 3$}
    \label{alg:jk_unroll}
\end{algorithm}

For larger $N$ values, \texttt{jk\_kernel} adopts a general implementation,
where the key difference is that the ERI digestion now occurs before the contraction.
This can be seen from Algorithm~\ref{alg:jk} that
the $\mathbf{J}$/$\mathbf{K}$ matrix is updated (e.g., line 23) within the loops over the basis functions.
Because the contracted ERIs no longer fit into registers or fast memory,
the contraction-then-digestion procedure will result in storing to and loading from
global memory, which may significantly hinder the performance.
Due to the same reason, the 2D integrals can only be stored in local memory.
They are computed once for all the Rys roots (line 5) to avoid increasing the number of updates to
the $\mathbf{J}$/$\mathbf{K}$ matrix.
Furthermore, to reduce global memory loads for retrieving the density matrix,
reusable strides (i.e., \texttt{P\_ac} and \texttt{P\_ad}) are cached in local memory,
potentially benefiting from optimal L1 and L2 cache utilization.
The same strategy applies to temporary stores of the potential matrix (i.e., \texttt{V\_ac} and \texttt{V\_ad}).
Additional scalar intermediates are also introduced (e.g., \texttt{P\_cd} and \texttt{V\_cd}),
which use registers for data loads and stores.
While shared memory (part of the L1 cache) could be used for data caching,
our experiment showed that it did not lead to better performance.
Since the cached intermediates are streaming rather than persistent data,
the compiler may optimize their memory usage more effectively than through manual manipulation.

Finally, it should be noted that the demand for local memory in Algorithm~\ref{alg:jk}
increases rapidly with the rise in angular momentum. For example, the integral class
of $\mathbf{(ii|ii)}$ requires $731$ KB of storage for the 2D integrals, which exceeds
the maximum allowed local memory of 512 KB per thread on the NVIDIA A100 GPU used here.
Therefore, our present implementation only supports ERIs with up to $g$ functions.

\begin{algorithm}
    get $\texttt{idx}[\ldots]$ \Comment*[h]{precomputed ERI to $I_x,I_y,I_z$ mapping}\;
    \For{$|\mathbf{ab}]$ \KwTo $|\mathbf{ab})$}{
        \For{$|\mathbf{cd}]$ \KwTo $|\mathbf{cd})$}{
            $t,w$ = \Rysroot{} \Comment*[h]{Rys roots/weights}\;
            $\mathbf{I}_x, \mathbf{I}_y, \mathbf{I}_z$ = \RR{$t, w$} \Comment*[h]{RRs for all roots}\;
            \Comment*[h]{$15$ is the size of a $g$ shell}\;
            allocate $\texttt{V\_ac}[15], \texttt{V\_ad}[15], \texttt{P\_ac}[15], \texttt{P\_ad}[15]$\;
            \For{$|d)$ \KwTo $|\mathbf{d})$}{
                $\texttt{V\_ad}[\ldots] = 0$\;
                $\texttt{P\_ad}[\ldots] = P_{(\mathbf{a}|d)}$\;
                \For{$|c)$ \KwTo $|\mathbf{c})$}{
                    $\texttt{V\_ac}[\ldots] = \texttt{V\_cd} = 0$\;
                    $\texttt{P\_ac}[\ldots] = P_{(\mathbf{a}|c)}$\;
                    $\texttt{P\_cd} =  P_{cd}$\;
                    \For{$|b)$ \KwTo $|\mathbf{b})$}{
                        $\texttt{V\_bc} =  \texttt{V\_bd} = 0$\;
                        $\texttt{P\_bc} =  P_{bc}$\;
                        $\texttt{P\_bd} = P_{bd}$\;
                        \For{$|a)$ \KwTo $|\mathbf{a})$}{
                            $\alpha,\beta,\gamma = \texttt{idx}[a,b,c,d]$\;
                            $g = 0$\;
                            \For{$i$ \KwTo \Range{$N$}}{
                                $g \mathrel{+}= \mathbf{I}_x[i,\alpha] * \mathbf{I}_y[i,\beta] * \mathbf{I}_z[i,\gamma]$\;
                            }
                            \AtomicAdd{$J_{ab}$, $g*\textup{\texttt{P\_cd}}$}\;
                            $\texttt{V\_ac}[a] \mathrel{+}= g * \texttt{P\_bd}$\;
                            $\texttt{V\_ad}[a] \mathrel{+}= g * \texttt{P\_bc}$\;
                            $\texttt{V\_bc} \mathrel{+}= g * \texttt{P\_ad}[a]$\;
                            $\texttt{V\_bd} \mathrel{+}= g * \texttt{P\_ac}[a]$\;
                            $\texttt{V\_cd} \mathrel{+}= g * P_{ab}$\;
                        }
                        \AtomicAdd{$K_{bc}$, \textup{\texttt{V\_bc}}}\;
                        \AtomicAdd{$K_{bd}$, \textup{\texttt{V\_bd}}}\;
                    }
                    \For{$|a)$ \KwTo $|\mathbf{a})$}{
                        \AtomicAdd{$K_{ac}$, \textup{\texttt{V\_ac}}$[a]$}\;
                    }
                    \AtomicAdd{$J_{cd}$, \textup{\texttt{V\_cd}}}\;
                }
                \For{$|a)$ \KwTo $|\mathbf{a})$}{
                    \AtomicAdd{$K_{ad}$, \textup{\texttt{V\_ad}}$[a]$}\;
                }
            }
        }
    }
    \caption{\texttt{jk\_kernel} for $N>3$}
    \label{alg:jk}
\end{algorithm}

\section{Nuclear gradient}\label{sec:grad}
The nuclear gradient of the electronic energy in the Hartree-Fock method is expressed as
\begin{equation}
    \bm{\nabla}_\mathbf{R} E = \bm{\nabla}_\mathbf{R} (E_{\text{core}} + E_J + E_K)
    - \sum_{ab} W_{ab} \bm{\nabla}_\mathbf{R} S_{ab} \;,
\end{equation}
where $E_{\text{core}}$ represents the energy associated with the one-electron core Hamiltonian,
$E_J$ and $E_K$ denote the Coulomb and exchange energies, respectively,
$\mathbf{W}$ is the orbital energy weighted density matrix,\cite{Pople1979}
and $\mathbf{S}$ is the overlap matrix.
In this work, only the computationally intensive Coulomb and exchange energy gradients
are evaluated on the GPU. The contributions from the one-electron integrals are
computed on the CPU
(with the exception of the nuclear-electron Coulomb attraction, which is computed on the GPU using a three-center integral code that is introduced in Ref.~\onlinecite{Wu2024}), allowing for concurrent execution with the GPU calculations.

A general $m$-th order ERI derivative using the Rys quadrature method can be straightforwardly evaluated  as follows,\cite{Flocke2008}
\begin{equation}
    \bm{\nabla}_\mathbf{R}^{m} \text{ERI} = \sum_{n} w_n \mathcal{F}_x [I_x(t_n)] \mathcal{F}_y [I_y(t_n)] \mathcal{F}_z [I_z(t_n)] \;,
\end{equation}
where $\mathcal{F}$ stands for a linear function.
Therefore, our GPU implementation of the nuclear gradients for the Coulomb and exchange energies
closely aligns with the implementation of \texttt{jk\_kernel} described in Section~\ref{sec:fock}.
For example, computing the nuclear gradient for
the integral class $\mathbf{(ss|ss)}$ is similar to
computing the energy for the integral class $\mathbf{(ps|ss)}$.

Nevertheless, the presence of the gradient operator adds an additional dimension
(i.e., the nuclear coordinates $\mathbf{R}$) to the ERIs.
It also increases the total angular momentum by one.
Moreover, while up to eight-fold permutation symmetry can be utilized in energy evaluation,
at most two-fold symmetry can be exploited for the ERI gradient.
Consequently, this leads to significantly higher register usage and memory footprint
for evaluating the RRs and the ERI gradient.
Additionally, more atomic operations are required if the Fock matrix gradient is to be computed and stored.

In order to overcome these bottlenecks, we double contract the ERI gradient with the density matrix
to directly obtain the energy gradient. Specifically, this involves computing the following
unique contributions on the fly:
\begin{align}
        &\bm{\nabla}_{\mathbf{R}_a} E_J = \sum_{abcd} (\bm{\nabla}_{\mathbf{R}_a}ab|cd) P_{ab} P_{cd} \;, \\
        &\bm{\nabla}_{\mathbf{R}_b} E_J = \sum_{abcd} (a\bm{\nabla}_{\mathbf{R}_b}b|cd) P_{ab} P_{cd} \;, \\
        &\bm{\nabla}_{\mathbf{R}_a} E_K = \sum_{abcd} (\bm{\nabla}_{\mathbf{R}_a}ab|cd) (P_{ac} P_{bd} + P_{ad} D_{bc}) \;, \\
        &\bm{\nabla}_{\mathbf{R}_b} E_K = \sum_{abcd} (a\bm{\nabla}_{\mathbf{R}_b}b|cd) (P_{ac} P_{bd} + P_{ad} D_{bc}) \;.
\end{align}
A general implementation of the GPU kernel (\texttt{ejk\_grad\_kernel}) for computing $E_J$ and $E_K$ gradients
is demonstrated in Algorithm~\ref{alg:grad} for $N>2$.
(Note here $N$ is determined after applying the gradient operator to the ERIs.)
Similar to Algorithm~\ref{alg:jk}, the 2D integral gradients are computed once for all Rys roots and stored in local memory (line 5). However, the gradients of $E_J$ and $E_K$
for the two nuclear centers $\mathbf{R}_a$ and $\mathbf{R}_b$ comprise only 12 scalar numbers,
which are cached in registers (line 1) and accumulated (lines 18--19) within the loops over the basis functions.
Notably, atomic operations are no longer required within these loops.
Instead, they are performed at the end of the kernel to write the results into global storage (lines 20--21),
totaling 12 operations for each shell quartet.
This can lead to significant performance gains compared to building the Fock matrix gradient.
Finally, for $N\leqslant2$, we cache the 2D integral gradients and other intermediates
in registers to minimize memory latency.

\begin{algorithm}
    $J_{\alpha\chi} = K_{\alpha\chi} = 0$ \Comment*[h]{$\alpha \in \{a,b\}$ $\chi \in \{x,y,z\}$}\;
    \For{$|\mathbf{ab}]$ \KwTo $|\mathbf{ab})$}{
        \For{$|\mathbf{cd}]$ \KwTo $|\mathbf{cd})$}{
            $t,w$ = \Rysroot{} \Comment*[h]{Rys roots/weights}\;
            $\mathbf{I}_{\alpha\chi}$ = \RRip{$t, w$} \Comment*[h]{gradient RRs}\;
            \For{$|d)$ \KwTo $|\mathbf{d})$}{
                \For{$|c)$ \KwTo $|\mathbf{c})$}{
                    $\texttt{P\_cd} =  P_{cd}$\;
                    \For{$|b)$ \KwTo $|\mathbf{b})$}{
                        $\texttt{P\_bc} =  P_{bc}$\;
                        $\texttt{P\_bd} = P_{bd}$\;
                        \For{$|a)$ \KwTo $|\mathbf{a})$}{
                            $\texttt{PJ} = P_{ab} * \texttt{P\_cd}$\;
                            $\texttt{PK} = P_{ac} * \texttt{P\_bd} + P_{ad} * \texttt{P\_bc}$\;
                             $g_{\alpha\chi} = 0$\;
                            \For{$i$ \KwTo \Range{$N$}}{
                                $g_{\alpha\chi} \mathrel{+}= \mathcal{F}_{\alpha\chi}(\mathbf{I}_{\alpha\chi}[i])
                                *\mathbf{I}_{\alpha\upsilon}[i]*\mathbf{I}_{\alpha\zeta}[i]$\;
                            }
                            $J_{\alpha\chi} \mathrel{+}= g_{\alpha\chi} * \texttt{PJ}$\;
                            $K_{\alpha\chi} \mathrel{+}= g_{\alpha\chi} * \texttt{PK}$\;
                        }
                    }
                }
            }
        }
    }
    \AtomicAdd{$E_J^{\prime}[\alpha, \chi]$, $J_{\alpha\chi}$}\;
    \AtomicAdd{$E_K^{\prime}[\alpha, \chi]$, $K_{\alpha\chi}$}\;
    \caption{\texttt{ejk\_grad\_kernel} for $N>2$}
    \label{alg:grad}
\end{algorithm}

\section{Results and discussions} \label{sec:Result}
In this section, we present the performance of our GPU-accelerated HF method,
implemented within the \textsc{GPU4PySCF} module.
All GPU calculations were performed on a single NVIDIA A100 GPU with 40 GB of VRAM.
For comparison, the CPU calculations were performed using the AMD EPYC 7763 CPUs with 32 threads.

First, we compare the wall times for restricted HF (RHF)
energy and nuclear gradient calculations using \textsc{GPU4PySCF}
with those of other GPU-accelerated HF codes,
including \textsc{GAMESS}\cite{Barca2020,Barca2021,gamessLatest} and
\textsc{QUICK}.\cite{QUICK,QUICKDFT}
Additionally, we provide results from the multi-threaded CPU code in \textsc{PySCF} as a reference.
The test set from Ref.~\onlinecite{Barca2021} was used,
which includes polyglycine ($\text{Gly}_n$) and RNA ($\text{RNA}_n$) molecules
at various sizes with 213--843 atoms and 131--1155 atoms, respectively,
using the STO-3G, 6-31G, and 6-31G(d) basis sets.
The integral threshold $\tau$ was set to $10^{-10}$.

We present the results in Table \ref{tab:timing}.
It is evident that \textsc{GPU4PySCF} outperforms \textsc{QUICK} in both
energy and nuclear gradient calculations, achieving speedups of over a factor of two.
For energy evaluations of the polyglycine systems, similar timings were observed
when comparing \textsc{GPU4PySCF} to \textsc{GAMESS}.
However, for the RNA systems, \textsc{GAMESS} outperforms \textsc{GPU4PySCF}, especially with the minimal basis set.
Furthermore, we also compare the computational scalings for different codes in Table~\ref{tab:scaling}.
Both \textsc{GPU4PySCF} and \textsc{QUICK} exhibit approximately quadratic scaling,
whereas \textsc{GAMESS} approaches linear scaling.
Finally, \textsc{GPU4PySCF} is one to two orders of magnitude more efficient than \textsc{PySCF}, highlighting its practical usefulness.

\begin{table*}
    \caption{
        Wall times (in seconds) for 10 self-consistent field (SCF) iterations and nuclear gradient calculations
        for various molecules and basis sets at the RHF level of theory.
    }\label{tab:timing}
    \scriptsize
    \centering
    \begin{tabular*}{1.0\textwidth}{@{\extracolsep{\fill}}llrccccccc}
        \hline\hline
        &&&\multicolumn{4}{c}{10 SCF iterations} & \multicolumn{3}{c}{Nuclear gradient}  \\
        \cline{4-7}\cline{8-10}
        System & Basis set & $N_\text{basis}$ & \textsc{GPU4PySCF} & \textsc{GAMESS}$^{a}$ & \textsc{QUICK} & \textsc{PySCF}& \textsc{GPU4PySCF} & \textsc{QUICK} & \textsc{PySCF}\\
        \hline
        \multirow{3}{*}{$\text{Gly}_{30}$} & STO-3G & 697 & 2.4 & 2.3 & 3.5 & 74.2 & 3.7 & 6.6 & 84.3 \\
        & 6-31G & 1273 & 6.4 & 15.2 & 12.4 & 238.9 & 7.8 & 18.2 & 288.9 \\
        & 6-31G(d) & 1878 & 17.4 & 34.4 & 44.1 & 477.4 & 29.1 & 61.2 & 579.6 \\
        [0.3em]
        \multirow{3}{*}{$\text{Gly}_{40}$} & STO-3G & 927 & 3.4 & 3.4 & 5.9 & 130.1 & 5.7 & 11.3 & 154.4 \\
        & 6-31G & 1693 & 10.4 & 19.2 & 23.0 & 430.1 & 12.4 & 31.1 & 576.6 \\
        & 6-31G(d) & 2498 & 28.9 & 45.9 & 80.2 & 880.7 & 49.7 & 107.0 & 1148.2 \\
        [0.3em]
        \multirow{3}{*}{$\text{Gly}_{50}$} & STO-3G & 1157 & 5.2 & 4.7 & 9.3 & 213.2 & 8.7 & 17.5 & 262.5 \\
        & 6-31G & 2113 & 16.0 & 24.3 & 38.0 & 686.8 & 19.8 & 49.2 & 1032.3 \\
        & 6-31G(d) & 3118 & 44.1 & 61.2 & 130.4 & 1444.4 & 74.8 & 168.1 & 2006.6 \\
        [0.3em]
        \multirow{3}{*}{$\text{Gly}_{60}$} & STO-3G & 1387 & 7.2 & 6.1 & 13.4 & 306.1 & 13.0 & 25.7 & 404.8 \\
        & 6-31G & 2533 & 21.3 & 31.1 & 57.6 & 1035.0 & 28.0 & 72.5 & 1688.3 \\
        & 6-31G(d) & 3738 & 61.2 & 80.8 & 190.9 & 2194.6 & 107.3 & 241.3 & 3278.1 \\
        [0.3em]
        \multirow{3}{*}{$\text{Gly}_{70}$} & STO-3G & 1617 & 9.4 & 8.0 & 18.5 & 421.7 & 17.3 & 35.8 & 614.5 \\
        & 6-31G & 2953 & 28.8 & 39.7 & 81.2 & 1439.1 & 37.6 & 101.4 & 2636.0 \\
        & 6-31G(d) & 4358 & 82.9 & 103.7 & 263.9 & 3145.9 & 144.3 & 343.8 & 5033.1 \\
        [0.3em]
        \multirow{3}{*}{$\text{Gly}_{80}$} & STO-3G & 1847 & 11.9 & 10.1 & 24.2 & 555.6 & 22.6 & 46.7 & 832.3 \\
        & 6-31G & 3373 & 35.9 & 48.4 & 109.0 & 1976.3 & 48.9 & 135.6 & 3959.9 \\
        & 6-31G(d) & 4978 & 107.0 &  & 352.4 & 4368.1 & 188.1 & 474.8 & 7461.2 \\
        [0.3em]
        \multirow{3}{*}{$\text{Gly}_{90}$} & STO-3G & 2077 & 15.0 & 12.5 & 31.3 & 713.0 & 28.5 & 58.9 & 1134.5 \\
        & 6-31G & 3793 & 45.0 & 58.5 & 140.7 & 2591.7 & 62.2 & 182.0 & 5750.5 \\
        & 6-31G(d) & 5598 & 134.8 &  & 554.3 & 5865.5 & 238.3 & 671.6 & 10723.6 \\
        [0.3em]
        \multirow{3}{*}{$\text{Gly}_{100}$} & STO-3G & 2307 & 18.7 & 14.9 & 39.7 & 908.5 & 40.4 & 75.5 & 1527.3 \\
        & 6-31G & 4213 & 56.3 & 71.4 & 213.1 & 3334.7 & 74.0 & 242.0 & 8093.5 \\
        & 6-31G(d) & 6218 & 172.8 &  & 668.1 & 7689.4 & 269.8 & 623.5 & 14977.0 \\
        [0.3em]
        \multirow{3}{*}{$\text{Gly}_{110}$} & STO-3G & 2537 & 21.8 & 17.9 & 49.4 & 1111.5 & 42.6 & 92.3 & 1992.5 \\
        & 6-31G & 4633 & 65.5 & 83.9 & 226.0 & 4217.0 & 92.7 & 283.6 & 11131.9 \\
        & 6-31G(d) & 6838 & 194.2 &  & 804.9 & 9977.5 & 326.5 & 1020.2 & 20468.9 \\
        [0.3em]
        \multirow{3}{*}{$\text{Gly}_{120}$} & STO-3G & 2767 & 26.0 & 20.8 & 58.5 & 1350.3 & 50.5 & 120.7 & 2569.5 \\
        & 6-31G & 5053 & 78.1 &  & 311.3 & 5295.4 & 111.0 & 375.4 & 15027.9 \\
        & 6-31G(d) & 7458 & 240.2 &  & 962.8 & 12702.1 & 427.3 & 1184.4 & 27402.3 \\
        [0.3em]
        \multirow{3}{*}{$\text{RNA}_1$} & STO-3G & 491 & 3.7 & 2.4 & 4.3 &  & 6.6 & 8.0 &  \\
        & 6-31G & 880 & 13.3 & 21.2 & 18.0 &  & 16.6 & 23.3 &  \\
        & 6-31G(d) & 1310 & 30.7 & 47.0 & 68.1 &  & 47.2 & 75.8 &  \\
        [0.3em]
        \multirow{3}{*}{$\text{RNA}_2$} & STO-3G & 975 & 15.7 & 7.0 & 21.1 &  & 28.3 & 33.7 &  \\
        & 6-31G & 1747 & 46.2 & 36.6 & 102.9 &  & 60.2 & 95.8 &  \\
        & 6-31G(d) & 2602 & 120.4 & 95.8 & 343.7 &  & 192.8 & 315.4 &  \\
        [0.3em]
        \multirow{3}{*}{$\text{RNA}_3$} & STO-3G & 1459 & 36.1 & 14.3 & 53.3 &  & 65.7 & 77.5 &  \\
        & 6-31G & 2614 & 98.9 & 64.5 & 263.8 &  & 130.2 & 219.7 &  \\
        & 6-31G(d) & 3894 & 318.2 & 184.2 & 880.0 &  & 508.9 & 717.2 &  \\
        [0.3em]
        \multirow{3}{*}{$\text{RNA}_4$} & STO-3G & 1943 & 67.1 & 24.5 & 101.8 &  & 124.2 & 140.4 &  \\
        & 6-31G & 3481 & 169.7 & 107.2 & 505.7 &  & 228.5 & 405.8 &  \\
        & 6-31G(d) & 5186 & 584.7 &  & 1625.5 &  & 865.8 & 1328.7 &  \\
        [0.3em]
        \multirow{3}{*}{$\text{RNA}_5$} & STO-3G & 2427 & 102.5 & 37.3 & 171.6 &  & 189.4 & 223.9 &  \\
        & 6-31G & 4348 & 268.6 & 166.3 & 1008.2 &  & 358.5 & 689.2 &  \\
        & 6-31G(d) & 6478 & 931.8 &  & 3162.4 &  & 1490.1 & 2451.7 &  \\
        [0.3em]
        \multirow{3}{*}{$\text{RNA}_6$} & STO-3G & 2911 & 150.7 & 53.9 & 267.5 &  & 291.2 & 342.5 &  \\
        & 6-31G & 5215 & 382.0 &  & 1454.6 &  & 514.5 & 1061.6 &  \\
        & 6-31G(d) & 7770 & 1317.3 &  & 4612.8 &  & 2172.0 & 3655.8 &  \\
        [0.3em]
        \multirow{3}{*}{$\text{RNA}_7$} & STO-3G & 3395 & 206.7 & 71.2 & 374.9 &  & 358.8 & 485.1 &  \\
        & 6-31G & 6082 & 509.8 &  & 2050.0 &  & 659.2 & 1430.6 &  \\
        & 6-31G(d) & 9062 & 1790.7 &  & 6258.1 &  & 2735.3 & 4680.6 &  \\
        [0.3em]
        \multirow{3}{*}{$\text{RNA}_8$} & STO-3G & 3879 & 283.0 & 97.2 & 480.2 &  & 515.8 & 683.8 &  \\
        & 6-31G & 6949 & 702.2 &  & 2845.2 &  & 893.2 & 2025.6 &  \\
        & 6-31G(d) & 10354 & 2513.8 &  & 8953.9 &  & 3793.4 & 7149.2 &  \\
        [0.3em]
        \multirow{3}{*}{$\text{RNA}_9$} & STO-3G & 4363 & 344.5 & 118.3 & 647.8 &  & 679.0 & 855.5 &  \\
        & 6-31G & 7816 & 837.0 &  & 3560.8 &  & 1163.6 & 2485.7 &  \\
        & 6-31G(d) & 11646 & 3008.3 &  & 11095.9 &  & 4624.3 & 8687.9 &   \\\hline\\[0.3em]
    \end{tabular*}
    \begin{flushleft}\footnotesize
        $^a$ Results are obtained from Ref.~\onlinecite{Barca2021}.
    \end{flushleft}
\end{table*}

\begin{table*}
    \caption{Observed computational scalings [$a$ in $O(N_{\text{basis}}^a)$] for energy and nuclear gradient calculations using different RHF codes.}
    \centering
    \begin{tabular*}{1.0\textwidth}{@{\extracolsep{\fill}}llccccccc}
        \hline\hline
        &&\multicolumn{4}{c}{10 SCF iterations} & \multicolumn{3}{c}{Nuclear Gradient}  \\
        \cline{3-6}\cline{7-9}
        system & basis set & \textsc{GPU4PySCF} & \textsc{GAMESS} & \textsc{QUICK} & \textsc{PySCF}& \textsc{GPU4PySCF} & \textsc{QUICK} & \textsc{PySCF}\\\hline
        \multirow{3}{*}{$\text{Gly}_{x}$} & STO-3G &  1.77 & 1.62 & 2.07 & 2.16 & 1.96 & 2.08 & 2.49 \\
        & 6-31G & 1.81 & 1.35 & 2.32 & 2.29 & 1.94 & 2.20 & 2.88 \\
        & 6-31G(d) & 1.91 & 1.31 & 2.28 & 2.42 & 1.91 & 2.15 & 2.81\\
        \multirow{3}{*}{$\text{RNA}_{x}$} & STO-3G & 2.08 & 1.81 & 2.29& & 2.10 & 2.14 \\
        & 6-31G & 1.91 & 1.28 & 2.43& & 1.94 & 2.16 \\
        & 6-31G(d) & 2.12 & 1.23 & 2.35& & 2.12 & 2.19 \\\hline
    \end{tabular*}
    \label{tab:scaling}
\end{table*}

\begin{figure}
    \centering
    \includegraphics[width=3.3in]{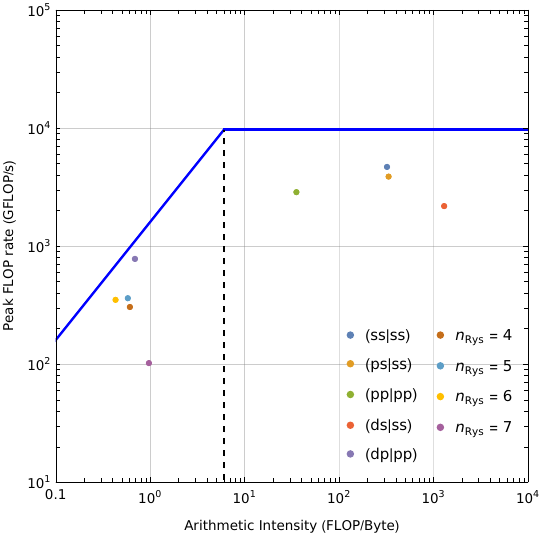}
    \caption{
        FLOP performance of the GPU kernels \texttt{jk\_kernel}
        analyzed using the roofline model on the NVIDIA A100 GPU.
        The solid blue line represents the official peak FP64 FLOP rate of $9.7$TFLOP/s (horizontal) and
        peak memory bandwidth of $1.6$TB/s (diagonal).
        The dashed black line indicates the machine balance of $6.1$FLOP/byte.
        The calculations were performed for a water cluster system consisting of 32 water molecules
        at the RHF/cc-pVTZ level of theory.}
    \label{fig:roofline}
\end{figure}

\begin{figure}
    \centering
    \includegraphics[width=3.3in]{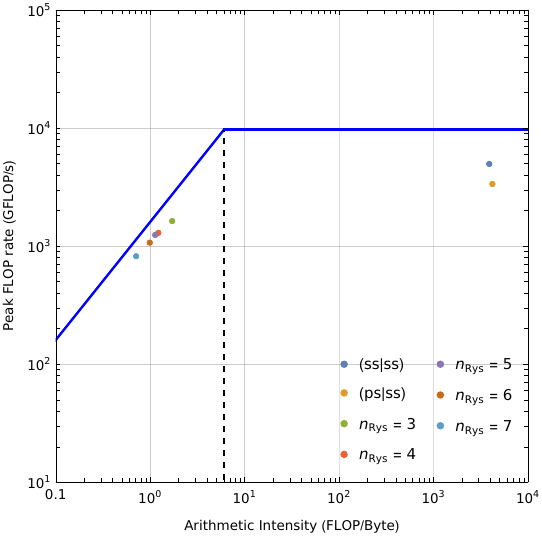}
    \caption{Same as Fig.~\ref{fig:roofline}, but for the GPU kernels \texttt{ejk\_grad\_kernel}.}
    \label{fig:roofline_grad}
\end{figure}

Next, we analyze the FLOP performance of the two GPU kernels
(i.e., \texttt{jk\_kernel} and \texttt{ejk\_grad\_kernel})
for various integral classes and $N$ values using the roofline model.
Profiling was performed on a water cluster system consisting of 32 water molecules,
using the cc-pVTZ basis set, which includes up to $f$ functions.
The results are displayed in Figs.~\ref{fig:roofline} and \ref{fig:roofline_grad}, respectively.

The roofline (solid blue line) represents the performance bound of the NVIDIA A100 GPU,
which includes a ceiling derived from the peak memory bandwidth (diagonal line)
and the processor's peak FLOP rate (horizontal line).
The dashed black line indicates the machine balance ($6.1$FLOP/byte).
Kernels with arithmetic intensity smaller than the machine balance are considered memory-bound,
while those with arithmetic intensity greater than it are compute-bound.

From Fig.~\ref{fig:roofline}, we observe that for most integral classes with $N\leqslant3$
[e.g., $\mathbf{(ss|ss)}$, $\mathbf{(ps|ss)}$, $\mathbf{(ds|ss)}$, and $\mathbf{(pp|pp)}$],
\texttt{jk\_kernel} is compute-bound and achieves an impressive FLOP rate
ranging from $2$TFLOP/s to $5$TFLOP/s.
However, there are exceptions, such as the integral class $\mathbf{(dp|pp)}$, which exhibits a
memory-bound character and limited FLOP performance.
This is due to the need for caching more than $234$ FP64 words for
intermediates per GPU thread, which exceeds the maximum number of registers (255 FP32 words) each thread can use by nearly a factor of two.
Consequently, intermediates that cannot fit into registers are likely stored in slow memory (known as register spilling),
resulting in significant memory latency when accessed frequently.

For $N>3$, \texttt{jk\_kernel} is always memory-bound due to the use of local memory for storing intermediates.
Nonetheless, the kernel generally utilizes the GPU hardware efficiently, as indicated by data points lying close to the roofline.
An exception is the integral class $\mathbf{(ff|ff)}$ (with $N=7$), which shows a potential loss of parallelization.
This is mainly because of the insufficient workload to fully occupy the streaming multiprocessors (SMs), as only O atoms contain $f$ shells, and each O atom contains only one shell of $f$ functions.

Similarly, \texttt{ejk\_grad\_kernel} shows a remarkable FLOP performance of over
$3$TFLOP/s for integral classes with $N\leqslant2$,
where intermediates can be cached in registers.
For $N>2$, the kernel is again memory-bound due to the use of local memory.
However, all data points in Fig.~\ref{fig:roofline_grad} lie close to the roofline,
indicating efficient utilization of the GPU hardware.
Notably, even for $N=7$, a FLOP rate of $0.8$TFLOP/s is achieved,
outperforming its \texttt{jk\_kernel} counterpart for Fock builds by a factor of eight.
This can be attributed to our integral direct approach as shown in Algorithm \ref{alg:grad}.
It eliminates the need to compute the Fock matrix gradient, which would otherwise be stored in global memory.
As as a result, significantly fewer atomic operations and slow memory accesses are performed, enhancing cache utilization.
In addition, the workload involved in gradient calculations is greater than that in Fock builds [e.g., the integral class $\mathbf{(ff|fd)}$ also corresponds
to $N=7$ when evaluating its gradient],
which keeps more GPU threads active and helps hide latency more effectively.

\section{Conclusions} \label{sec:Conclusion}
In this work, we introduced the \textsc{GPU4PySCF} module, and in particular, the core ERI CUDA kernels that form the starting point for accelerating quantum chemistry calculations. As an example of their use, we described a GPU-accelerated HF method for energy and nuclear gradient calculations, including the detailed optimizations required to achieve high GPU efficiency.

The GPU acceleration of quantum chemistry is integral not only to advancing traditional quantum chemistry calculations, but also to bringing
quantum chemical methods and data into new disciplines such as machine learning.  We hope that by providing a community-based, open-source implementation of GPU accelerated quantum chemistry algorithms, we can help the growth of quantum chemistry in these areas.
Indeed, we note that as a living open-source code, at the time of writing \textsc{GPU4PySCF} already contains new contributions targeted at these directions.\cite{Wu2024}

\section*{Acknowledgements}
We acknowledge the generous contributions of the open-source community to the \textsc{GPU4PySCF} module.
RL and QS contributed equally to this work.
Work carried out by QS (development of initial ERI code and Fock build) was performed as a part of a software contract with GKC through the California Institute of Technology, funded by internal funds. RL (development of gradient ERIs and gradient code) and GKC (project supervision) were supported by the US Department of Energy, Office of Science, through Award No. DE-SC0023318. XZ (additional data analysis) was supported by the Center for Molecular Magnetic Quantum Materials, an Energy Frontier Research Center funded by the U.S. Department of Energy, Office of Science, Basic Energy Sciences under Award No. DE-SC0019330.
This research used resources of the National Energy Research Scientific Computing Center (NERSC), a U.S. Department of Energy Office of Science User Facility located at Lawrence Berkeley National Laboratory, operated under Contract No. DE-AC02-05CH11231 using NERSC award ERCAP-0024087.

\section*{References}
\bibliography{references}
\end{document}